\documentclass[aps,prb,twocolumn,groupedaddress]{revtex4}
\usepackage{graphics,epsfig}
\begin{document}

\title{Density-functional study of adsorption of isocyanides on the gold (111) surface.}

\author{Yulia Gilman}
\affiliation{Department of Physics and Astronomy, State University of New York, 
Stony Brook, NY 11794-3800}

\author{Philip B. Allen}
\affiliation{Department of Physics and Astronomy, State University of New York, 
Stony Brook, NY 11794-3800\footnote{Permanent address.}, 
\\and Center for Functional Nanomaterials, Brookhaven National Laboratory, Upton, New York 11973-5000}

\author{Mark S. Hybertsen}
\affiliation{Department of Applied Physics and Applied Mathematics, 
and Center for Electron Transport in Molecular Nanostructures, 
Columbia University, New York, NY 10027}

\begin{abstract}

Density functional theory within the generalized-gradient approximation 
is used to study the adsorption of 
the isocyanides CNH and CNCH$_3$ on the gold (111) surface at several coverages.
It is found that these molecules are highly selective 
in their adsorption site preference.
Adsorption is possible only at the top site, 
although binding is rather weak and dipole-dipole repulsion prevents
binding at coverages of 1/3 ML and higher. 
At all other high symmetry sites considered (hcp, fcc, bridge),
the isonitriles are not bound.
To reveal the main mechanisms of bonding, 
the isonitriles are compared to CO and ammonia using 
Au-X radicals as model systems.
The systematic trends are understood in terms of $\sigma$ donation
from the ligand lone pair and $\pi^*$ back donation.
Finally, CNH is found to be more strongly bound to an
undercoordinated gold atom (adatom) on the surface.
\end{abstract}

\maketitle

\section{Introduction}

Gold is widely chosen as an electrode material for studies of molecular conductance 
for a number of practical reasons, including ease of deposition and patterning and 
presentation of a surface that is relatively easy to clean and maintain.
Another important factor has been the extensive base of experience in forming 
self assembled monolayers (SAMs) of organic molecules, linked to the gold surface via
a sulfur atom.
The initial discovery of the thiol-gold route to self assembly \cite{nuzzo}
led to rapid growth in the study of the formation and properties
of organic layers \cite{ulman,schreiber04}.
The S-Au bond is relatively strong ($\sim$1.7 eV), but the barrier to lateral
motion between adsorption sites is rather low ($\sim$0.1 eV), facilitating formation of
ordered monolayers \cite{ulman}.
Sulfur has been alsmost exclusively used as the ``alligator clip'' \cite{tour}
in the experiments probing conductance at the molecular scale \cite{reed,salomon}.
However, the substantial variation in the measured conductance 
for basic molecules such as alkanes contacted using the Au-S link,
between different groups and different measurement techniques, remains a puzzle.
One possibility is sample to sample variation in the local structure
of the Au-S link \cite{basch}.
This is consistent with the low barriers for motion that facilitate assembly
and the known mobility of Au atoms 
during formation of the thiol linked layer \cite{poirrer,schreiber00}.
Also, the atomic scale structure of the interface of assembled layers with Au-S linkages
has proved to be a very delicate problem for {\it ab-initio} methods \cite{schreiber00},
although a recent study appears to reconcile extensive 
data for decanethiols \cite{fischer}.
Overall, despite wide usage, the Au-S link displays a number of critical issues for 
application in molecular conductance studies.
Recent research has been directed to find alternative link chemistries, 
e.g. Ru or Mo carbene \cite{nuckolls,mcbreen}.

An alternative link chemistry for a gold electrode that shows more 
selectivity in the local bonding geometry is an attractive possibility in
the context of molecular scale conductance.
Although Au surfaces are relatively unreactive,
other candidates have been explored.
A very recent study demonstrated the utility of amine terminated molecules
for molecular conductance studies \cite{venkataraman}.  The Au-NH$_2$-R links
proved to be flexible and reproducible.
Another example, with a similar bonding character, is isocyanide (CN-R, e.g. CNH), 
a less-stable isomer of cyanide (NC-R, e.g. NCH).
Well-established synthetic methods exist for the CN- linker 
with various molecules (R). Bonding to Pt or Au electrodes has been observed \cite{lin}.
In contrast to the Au-S link, 
adsorption of isocyanide (Au-CN) linked molecules on gold surfaces is
much less studied.
Most experimental works on this subject consider adsorption of the molecules
on gold nanoparticles \cite{8,9,10,11}.  
Henderson {\it et al.} \cite{7} studied the formation of self-assembled monolayers of 
diisocyanides on the gold (111) surface.
It was concluded from reflection absorption infrared (RAIR) spectra 
that the molecules with rigid chains 
between two CN- groups (such as biphenyldiisocyanide) attach 
to the gold surface through only one isocyanide group while the other group 
remains free.
Ellipsometry data indicate that in these cases, the molecular axis is 
perpendicular to the surface.
Molecules with flexible alkyl chains can attach to the gold surface 
through both isocyanide groups, which makes them unlikely 
candidates for molecular conductance studies. 
The preferred adsorption site and geometry 
remain as unresolved questions in the literature.
Angelici {\it et al.} observed the peak in the infrared spectra 
associated with N-C stretch mode to be shifted 50-70 cm$^{-1}$ 
higher upon adsorption on gold powder \cite{8,9}.   
They concluded that isocyanides bind to the surface at the top site. 
Henderson {\it et al.} \cite{7} observed a similar blue shift 
for adsorption on the gold (111) surface, but concluded 
that both atop and three-fold adsorption configurations are possible.

The earliest of the few theoretical works on the adsorption of isocyanide 
on gold considered phenyl isocyanide 
attached to a single gold atom, a gold dimer and a 
trimer \cite{13}.  
The calculations showed that for each gold cluster, the binding was to a single Au atom.
Bonding motifs analogous to the bridge and hollow sites on the (111) surface
were not observed.
The results suggest site selectivity for the isocyanide link, although
the details of the adsorption energy and the geometry on the Au(111)
surface remain to be determined.
Recently several groups \cite{1,2} successfully fabricated self-assembled monolayers 
of various diisocyanide molecules 
sandwiched between two gold electrodes, and studied electrical transport 
through them.  
The impact of the isocyanide link on the conduction has received 
some attention from theorists as well.  
Conductance and voltage-current characteristics of molecules 
connected to metallic electrodes were calculated using combined
density functional theory (DFT) and non-equilibrium Green's function 
methods \cite{3}.
However, the adsorption geometry was not explicitly determined
by energy minimization.

This paper presents a study of the adsorption of isocyanide linked
molecules on the gold (111) surface using a density functional theory (DFT) approach.  
We focus on the binding of the essential link element to the metal surface,
considering the simplest case, a CNH molecule.
Calculations performed for CNCH$_3$ confirm the results. 
The basic electronic structure of the CNH molecule suggests that the filled
lone pair on the C atom will be available for donor-acceptor type bonding
to Au complexes, in analogy with ammonia (NH$_3$) and CO.
We chose to draw a comparison between CNH, CO and NH$_3$ 
in order to better understand the factors that affect adsorption.
The case of CO has been extensively studied 
for many transition metal surfaces \cite{trends}.
There are also recent calculations for the case of NH$_3$ on Au(111) \cite{ammonia}.
The standard Blyholder picture \cite{Blyholder} describes the bonding
in terms of the trade-off between $\sigma$ donation from the lone pair to the
metal and back donation from the metal to the empty $\pi^*$ molecular orbital.
Two main themes emerge: (1) the interplay of the electronic couplings
in the Blyholder picture with
the corresponding charge transfers; (2) the impact of the dipole moment, both 
that of the molecule prior to adsorption and the induced dipole upon adsorption.
In the next section, the methodology and the basic results of the DFT
calculations are described.
The physical interpretation of the results is discussed in Sect. III.

\section{Methodology and results}

The adsorption of the target molecules on the Au surface is studied 
using a computational approach based on DFT.
The main calculations are done using the generalized gradient approximation (GGA)
of Perdew, Burke and Ernzerhof (PBE) \cite{PBE}. 
Different forms of GGA are widely used for metal
surface and adsorption problems in order to address the
large errors in adsorption energy found 
with the local density approximation (LDA),
although the accuracy still varies with the system 
under study \cite{gross,PBEtest,hammer,trends,feibelman}.
For comparison, selected calculations are also done 
with the LDA \cite{LDA,goedecker}.
To have an initial picture of the bonding, the target molecules bonded to a
single Au atom are considered 
using the NRLMOL DFT code \cite{NRLMOL}.
For the light atoms (C, N, and H), an all electron basis is used.
A slight modification \cite{NRLMOLpsp} of norm-conserving 
Troullier-Martins pseudopotentials is used for gold atoms.
NRLMOL uses Gaussian basis sets optimized for 
density-functional calculations \cite{NRLMOLbasis}. 
The exponents and contraction coefficients are determined by 
optimizing the total energy of the free atom in the ground state.
For the valence states, the number of independent contractions used
(s, p, d) for each atom is as follows: H (4, 3, 1), C (5, 4, 3), N (5, 4, 3)
and Au (4, 2, 4).
It has been shown that these basis sets are well-converged and 
that they have negligible basis set superposition error. 
Further details of the basis sets can be found in Ref. \cite{NRLMOLbasis}.
Radicals were treated with spin-polarized, unrestricted calculations.

The adsorption on the surface was modeled with the standard scheme
of a periodically repeated slab geometry.  
We found that the slab of four layers of gold atoms was enough to 
reproduce key adsorption characteristics such as the binding energy.
The molecules are adsorbed on one side of the slab.
For most of the calculations, the molecular structure is relaxed, 
with the molecular axis kept perpendicular to the surface and the Au atoms in
the slab frozen at their bulk positions.
Relaxation is carried out until the
maximum force is less than 0.15 eV/$\AA$.
For selected cases, the top layer of Au atoms was allowed to relax.
We also explored bonding of CNH to an Au adatom in the hcp hollow site,
allowing the adatom and the surface Au atoms to relax.
Most of these calculations were done using
the WIEN2k DFT code (full potential and linearized 
augmented plane waves \cite{12}).
For comparison, some calculations were also done with the 
ABINIT package \cite{abinit,gonze},
which employs a plane wave basis set and pseudopotentials.
LDA calculations were done with 
Hartwigsen-Goedecker-Hutter (HGH) pseudopotentials \cite{HGH},
while GGA calculations were done with Troullier-Martins type pseudopotentials \cite{TM,FHI}.
It was checked that the energy is converged with respect to the number of 
k-points as well as the size 
of the basis set.
To facilitate analysis, the partial densities of state (PDOS) were plotted in selected
cases.
In the WIEN2K code, a modified tetrahedron method \cite{tetrahedron} 
is used for density of state calculations;
the PDOSs are calculated by projecting the wave functions 
onto the spherical harmonic basis functions centered on the atoms in question
inside a sphere centered on the atom 
(atomic sphere radii are 1.06 $\AA$ for Au atoms, 0.58 $\AA$ for C and N, and 0.29 $\AA$ for H). 

We start by considering binding of the CNH and CNCH$_3$ molecules to a single gold atom.
The binding energy of the Au atom, the gold to ligand distances, and 
the dipole moments of the free molecules and 
the molecules bound to the gold atom are reported in Table I.
The results for CNH and CNCH$_3$ are very similar.
For comparison, results for CO and NH$_3$ are also shown.
The first step of relaxation allows only configurations with axial symmetry. 
The axial CO and CNH molecules are very similar in their electronic structure. 
The highest occupied molecular orbital (HOMO) is the lone pair on the C
and there is a doubly degenerate $\pi$ bond between C and the N or O. 
For NH$_3$, the N lone pair is the HOMO, but there are no $\pi$ bonds.
The $\sigma$ part of the interaction in all three is due to coupling of the
lone pair orbital on the ligand to the $d_{z^2-r^2}$ and s states on the Au.
This results in partial transfer of electron density onto the Au atom.
Because the Au s state is half filled, the frontier, anti-bonding $\sigma$ state
in the complex is also half filled.  
The initial ligand lone pair orbital energy is highest
for NH$_3$, comparable to the Au s state, 
and successively lower for CNCH$_3$, CNH and lowest for CO.  

The $\sigma$ donation is partially balanced by the back donation 
from the Au $d_{zx}$ and $d_{yz}$
into the empty $\pi^*$ doublet on the molecule for CO, CNH and CNCH$_3$.
This balance is slightly different for AuCNH and AuCNCH$_3$ relative to AuCO.  
In the AuCO case, the C-O bond length increases by 0.004 $\AA$ relative to CO.
In the AuCNH (AuCNCH$_3$) case, 
the C-N bond length decreases by 0.007 $\AA$ relative to CNH (CNCH$_3$).
This difference suggests less $\pi^*$ back donation in the AuCNH case.
The reduced back donation is consistent with the $\pi^*$ doublet being about
1 eV higher in CNH and even a bit higher
for CNCH$_3$.  This is also consistent with the trend of increasing Au-C
bond length from AuCO to AuCNH to CNCH$_3$. 
The trend for net charge transfer to the Au as measured by Mulliken type analysis
is consistent with larger $\sigma$ donation and smaller $\pi$ back donation
from AuCNCH$_3$ to AuCNH to AuCO.
The binding energy follows the same order, but the impact 
of larger $\sigma$ donation and smaller $\pi^*$ back donation are competitive,
with the $\sigma$ donation dominating the trend.
By comparison, the NH$_3$ molecule shows 
the strongest $\sigma$ donation, with the largest charge transfer to Au,
but offers no empty $\pi$ space for back donation.  
Its binding energy to a gold atom 
is 0.60 eV, less than for CO. 
Correspondingly the Au-N bond length is substantially larger (2.277 $\AA$). 
The role of the $\pi$ states is further highlighted by the
results of full relaxation, which leads to bent structures for 
Au-CNH, Au-CNCH$_3$ and Au-CO. 
This bending is due to the pseudo Jahn-Teller effect. 
The lowered symmetry allows mixing between the half occupied $\sigma$ antibonding level 
in the complexes and one of the empty $\pi^*$ states.
The resultant energy gain depends on the gap between $\sigma$ and $\pi^*$ states.
It is largest for the Au-CO complex and smallest for Au-CNCH$_3$; 
the Au-CO complex has the
smallest gap between $\sigma$ and $\pi^*$ states, 
while Au-CNCH$_3$ has the largest. 

In comparison to CO, 
a key difference is that CNH and NH$_3$ have a much larger dipole moment.
Table 1 shows the calculated dipole moments which agree well with the
measured values.
In all three cases,
there is a substantial increase in the dipole of the gold-ligand complex relative
to the isolated ligand.
The dipole moment of the AuCO complex is 1.6 D for the linear configuration, 
compared to the 
small dipole of the CO molecule.
The dipole of the AuCNH complex increases by 2.4 D over the CNH ligand,
while the increase is 3.2 D for the NH$_3$ case.
The increase in dipole can not be simply seen in a point charge approximation
with associated charge transfer (e.g. from a Mulliken type analysis).
Instead, the extension of the lone pair electrons at the C end of
the ligand onto the Au by quantum mechanical mixing with the Au s state is responsible,
partly balanced by the $\pi^*$ back donation.
The reduced back donation in the AuCNH case results in a larger change in dipole.
In the NH$_3$ case there is no back donation and the dipole change is even larger.
On the other hand, the distortion to the bent form in the Au-CNH
and Au-CO cases results in a much reduced dipole,
consistent with the enhanced back donation from the Au to the C.

In addition to the hybridization considerations, the dipole also influences
the relative binding of Au to the ligands.
There is an additional attraction in the Au-CNH complex due
to polarization of the Au by the dipole.
This influences the differences in binding energy, in addition to the systematic
hybridization differences.
To estimate this contribution, we use the calculated polarizability of the
Au atom \cite{ammonia} together with a point charge model of the molecule
derived from fitting to the self consistent electrostatic potential and the
calculated dipole moment.
For the case of AuCNH, we estimate that polarization contributes 0.1 eV to
the binding energy. For AuNH$_3$, the effect is even smaller (0.01 eV).

So far as we know, there are no direct measurements of the Au binding energy
in these radicals.
The Au-NH$_3$ radical was studied using both DFT approaches and quantum chemistry
approaches for evaluation of
the correlation energy (e.g. coupled cluster with singles, doubles 
and triple corrections, CCSD(T)) \cite{lambropoulos}.
Substantial care was also taken to account for basis set superposition
errors, which can be a significant issue for these correlation techniques 
in particular \cite{dargel}.
The Au binding energy in this case was deduced to be 0.78$\pm$0.06 eV,
similar to the GGA-PBE value we obtain.
There are several studies of the Au-CO radical with DFT and quantum chemistry
approaches \cite{schwerdtfeger, liang, mendizabal, wu}.
Our result for the lowest energy
geometry, the bent form, agrees reasonably well 
with previous values based on DFT with GGA: 
0.71 eV \cite{liang} and 0.80 eV \cite{wu}.
However, the correlated electron calculations consistently give smaller
binding energies, ranging from 0.14 to 0.56 eV depending on the details
\cite{schwerdtfeger, mendizabal}.
Given concerns about basis set superposition errors, it is premature to draw
a final conclusion from the quantum chemistry calculations in this case.

\begin{table*}

\caption{Comparison of Au-X complexes, where X is CO, NH$_3$, CNH and CNCH$_3$. All characteristics are calculated with NRLMOL DFT 
code using both PBE or LDA exchange-correlation potentials. Experimental values of the dipole moments are given in parentheses when available.
The X dipole column corresponds to the dipole moment of the isolated molecule X in the case of linear configurations. 
In the case of a bent configuration, it corresponds to the dipole moment of the ligand X 
with the geomtery taken from the fully relaxed bent Au-X complex.}
\begin{tabular}{lcccccc}
\hline
\hline
Molecule X & Au-X geometry &  Functional & Binding energy, eV & Au-X bond length, $\AA$ & X dipole, D & Au-X dipole, D\\
\hline
CO & linear &   PBE & $0.68 $  &  $1.997$ & $0.20 (0.12)$\footnote{J. S. Muenter, J. Mol. Spectrosc. {\bf 55}, 490 (1975).
} & $1.65$ \\
CO & linear &   LDA & $1.27  $  &   $1.948$ & $0.23$  & $1.57$ \\
CNH & linear &   PBE & $0.93  $  &   $2.002$ & $3.09 (3.05)$\footnote{G.L. Blackman, R.D. Brown, P.D. Godrey, and H.I. Gunn, Nature {\bf 261}, 395 (1976).}  & $5.46$ \\
CNH & linear &   LDA & $1.54  $  &  $1.957$ & $3.20$  & $5.50$ \\
NH$_3$ & linear &   PBE & $0.60  $  &   $2.277$ & $1.53$  & $4.72$ \\
NH$_3$ & linear &   LDA & $1.07  $  &   $2.188$ & $1.54$  & $4.89$ \\
CNCH$_3$& linear    & PBE & $ 0.99  $  &   $2.011$ & $3.95 (3.83)$\footnote{S.N. Ghosh, R. Trambarulo, and W. Gordy, J. Chem. Phys. {\bf 21}, 308 (1953). 
}  & $6.97$ \\
CO & bent &   PBE & $0.87 $  &  $2.005$ & $0.14$ & $0.35$ \\
CNH & bent &   PBE & $1.04  $  &   $1.989$ & $2.69$  & $2.89$ \\
CNCH$_3$ & bent   & PBE & $ 1.05  $  &   $2.019$ & $3.94$  & $5.74$ \\
\hline
\hline

\end{tabular}
\end{table*}

\begin{figure}
\includegraphics{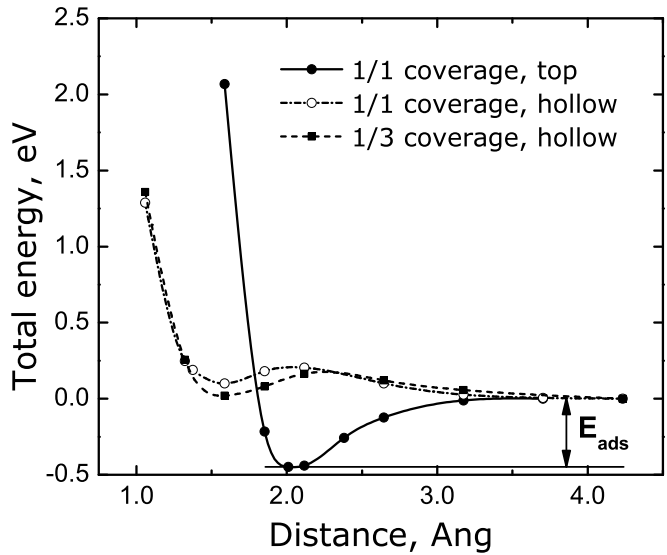}
\caption{Total energy (GGA-PBE) per unit cell for HNC monolayers with $1/1$ and $1/3$ coverages on the gold slab { \it vs.} distance $d$
to the surface. Molecules are at the top or filled hollow site with molecular axis perpendicular to the surface.  The indicated adsorption energy is relative to a uniform molecular layer well separated from the surface.
}
\end{figure}  

Turning to the adsorption on the surface,
we initially consider a full monolayer of CNH
adsorbed on one side of a slab of four gold layers.
The supercell contains one molecule and four gold atoms, two of them being 
surface atoms. 
First, the molecules are placed at the top site (above the surface Au atom) 
and the distance from the carbon atom to the surface is varied.  The molecular axis is kept perpendicular to the 
surface. The observed 
dependence of total energy  { \it vs.} distance $d$ is shown in Fig.1.
The binding energy $E_{ads}$ is the energy per molecule required to separate
the layer of CNH from the Au (111) surface, leaving the free standing
CNH layer otherwise unchanged.  
The value of $E_{ads}$ is found to be 0.43 eV and the optimal distance is 2.01 $\AA$. 
In order to check that four layers of gold are enough to represent 
the surface, we repeated the calculation for three, five, and six 
gold layers in the slab.  The results value of $E_{ads}$ varies
from 0.453 to 0.431 eV, 
indicating that the choice of four layers is justified. 

When a molecule is placed at the filled (hcp) hollow site, the
energy {\it vs.} distance (see Fig. 1) is
qualitatively different from that at the top site.  Although there is a
local minimum at around 1.6 $\AA$, the energy 
is minimal when the molecule is far away from the surface. 
This indicates that the CNH molecule cannot be stabilized
at the hollow hcp site.  The same result is obtained for the empty hollow (fcc) 
and for the bridge sites.

The next candidate for a stable configuration at uniform commensurate coverage
is a monolayer with one molecule per three surface atoms, forming a 
$\sqrt{3}\times\sqrt{3}R30^{\circ}$ equilateral triangular lattice on the surface, 
where R=2.883 $\AA$ is the distance between nearest neighbor gold atoms. 
The distance between molecules in this case is 4.944 $\AA$.
For the CNH molecules positioned at the top site, we relax the monolayer keeping the molecular axis normal 
to the surface and gold atoms frozen.
The adsorption energy is found to be 0.18 eV relative to the separated slab-monolayer system.
For the hcp site and $\sqrt{3}\times\sqrt{3}R30^{\circ}$ surface unit cell, similar to the $1\times 1$ case, 
the energy is minimal 
when molecules are far away from the surface, although there is a local minimum in energy curve (see Fig. 1).
This shows that adsorption is not possible at the hcp site.
Similar results are obtained for the CNH monolayer with 1/4 coverage ($2\times 2$ surface unit cell) and 
the CNCH$_3$ monolayer with 1/3 coverage 
(see Table II).

\begin{table}
\caption{Stabilization energies from WIEN2K calculations for various 
molecules obtained with GGA-PBE exchange-correlation potential.}
\begin{tabular}{lccc}
\hline
\hline
Molecule & Coverage & Site & Stabilization energy, eV \\
\hline
CNH & 1/1 & top &  $0.43 $ \\
CNH & 1/1 & hcp &   no adsorption \\
CNH & 1/1 & fcc &   no adsorption \\
CNH & 1/1 & bridge &   no adsorption \\
CNH & 1/3 & top &   $0.18 $ \\
CNH & 1/3 & hcp &   no adsorption \\
CNH & 1/4 & top &   $0.20 $ \\
CNH & 1/4 & hcp &   no adsorption  \\
CNCH$_3$ & 1/3 & top &    $0.22 $ \\
CNCH$_3$  & 1/3 & hcp &   no adsorption  \\
CO & 1/3 & top &   $0.20 $ \\
\hline
\hline
\end{tabular}
\end{table}

So far binding of pre-formed monolayers was considered. 
For adsorption from the gas phase, the energy cost of forming the free monolayer 
must be taken into account. 
The CNH molecule has a large dipole moment, as noted above.
In the case of 1 ML coverage the distance between molecules in the 
monolayer is 2.883 $\AA$.
The point dipole approximation gives a crude estimate of about 1.4 eV per 
molecule for the energy cost of monolayer formation. 
This is much larger than the energy gained upon adsorption. Therefore HNC molecules 
cannot adsorb on the (111) gold surface at 1 ML coverage.
For smaller coverages (1/3 and 1/4 ML) the distance between molecules is large enough for the point 
dipole approximation to be accurate.
We calculate an energy cost of 0.26 eV per molecule to form a 
CNH monolayer at 1/3 ML coverage and 0.17 eV at 1/4 ML coverage. 
When compared to the adsorption energies in Table II,
the formation of 1/3 ML is endothermic, while the 1/4 ML coverage is
slightly exothermic, about 0.03 eV per molecule.

Just as noted for the AuCNH complex, adsorption leads to a change in dipole moment, 
due to both the charge transfer and the polarization
of the molecules and the gold surface. 
For sparse coverages (less than 1/3), the induced dipole should not depend much on the
distance between molecules. 
This extra dipole gives a coverage dependent contribution which is included in the 
adsorption energy reported in Table II.
The scaling of this term is inversely 
proportional to the cube of the intermolecular distance. 
Using this approach and the data in Table II, 
we extrapolate the adsorption energy in the dilute limit to be 0.23 eV 
and the effective dipole moment of the adsorbed monolayer to be 3.35 D.
Thus the induced dipole moment in the monolayer is about 0.3 D, a considerably
smaller change than found for the AuCNH complex.
We also estimate the contribution to the binding that derives from
the classical polarization response of the metal surface to the 
initial dipole on the CNH.
We use the point charge model mentioned previously for CNH
and assume a classical image potential to arrive at about 0.1 eV. 
In comparison to the relatively small extrapolated net adsorption energy,
polarization plays a noticeable role.

Given the extra repulsion from the induced dipole,
one might have expected the CNH to adopt a bent configuration on the surface.
In the radical, this dramatically reduces the dipole, as seen in Table I.
This has been investigated by considering small distortions of the CNH
away from the vertical a-top structure at 1/3 ML coverage (molecular axis was tilted 
by 10$^{\circ}$, while the geometry of the molecule was otherwise unchanged).
Relaxation of this initial configuration results in an essentially vertical structure.
Alternatively, we have started from a bent CNH configuration at 1/4 ML coverage
(with Au-C bond vertical, CNH molecule positioned as in the bent Au-CNH complex) 
and relaxed the structure.
We find a smooth path back towards the vertical geometry.
Evidently the back donation driving force, characterized further in the next section,
is weaker in the adsorbed case than in the radical.

In the neutral radical, the net effect of the donor-acceptor bonding is relatively weak
due to the partially occupied Au s-orbital.  
For an Au cation, or a complex which partially withdraws the Au s-electron, the 
binding can be substantially stronger.
Experiments and calculations have been analyzed for a series of
ligands bonding to the gold cation \cite{schroder}.
The estimated bond energies for Au$^+$CO, Au$^+$NH$_3$, and Au$^+$CNCH$_3$
are 2.1, 3.1 and 3.1 eV respectively.
These are all substantially larger than for the corresponding neutral radicals.
Furthermore, it is known that isocyanides as  well as amines and phosphines passivate
gold nanoparticles \cite{8, aminecapped,phosphinecapped}.
This raises the interesting question of the character of CNH bonding to
undercoordinated gold atoms on the surface.
To probe this question, we calculated the binding of CNH to an Au adatom
using ABINIT.
The adatom was relaxed, together with the first full surface layer,
in the hcp hollow site of a 4 ML slab with 1/4 coverage.
Then the CNH was added to the system and fully relaxed.  
The final geometry is close to vertical with a reduced Au-C bond
length (1.958 $\AA$) and slightly longer C-N bond length (by $<$0.01 $\AA$)
relative to the original a-top geometry.
The binding energy 
is increased by 0.9 eV, relative to the a-top adsorption site on the flat surface.
This binding energy refers to an array of CNH molecules in the same
unit cell relative to the gold surface including the adatoms.
The formation energy of the adatom is not explicitly included
since we consider the adatom to be a proxy for undercoordinated gold 
atoms on the surface of a nanoparticle or other rough gold surfaces.

Finally, it is well known that the DFT method using 
the LDA overestimates binding energies. 
However it is usually expected that 
general trends are reproduced correctly. 
Some calculations for the radicals were repeated with 
the LDA exchange-correlation functional (Table 1). 
As expected, LDA systematically overestimates binding energies by about 0.6 eV and
underestimates the Au-ligand bond lengths.
Also, some of the slab calculations of CNH adsorption on Au(111) 
were repeated using LDA.
Adsorption energies calculated with ABINIT for a 1/4 ML coverage
are 0.95 eV, 1.09 eV and 1.12 eV for the top, hcp and fcc sites respectively.
Similar results are obtained with WIEN2k code for 1/3 ML coverage: 
0.80 and 1.03 eV for the top and hcp sites. 
There is a qualitative difference between the GGA-PBE and LDA results, the GGA being more realistic where experimental
tests are available. 
With GGA the top site is preferred and there is no adsorption at the hcp or fcc sites. 
With LDA, on the contrary, adsorption at the hcp site is possible 
and it is energetically favored over adsorption at the top site.

\section{Discussion}

We have found that isocyanides selectively bind to the atop site on the flat
gold (111) surface, with a small adsorption energy.
The C-N bond in the isocyanide is vertical.
This binding geometry is consistent with experiments described in the introduction:
top site adsorption is observed \cite{8,9} and ellipsometry data 
indicate vertical orientation of adsorbed isocyanides \cite{7}.
At the present there are no measurements of binding energies 
or saturation coverage of isocyanide molecules on the gold surface
for direct comparison to our calculated adsorption energy. 
The CO molecule, on the other hand, is very well studied.
As noted, its electronic structure is similar to that of the CNH molecule.
We performed calculations identical to those described above for 
CO on the gold (111) surface and obtained 0.2 eV for the adsorption energy,
in good agreement with previous calculations,
but less than the experimental value 0.4 eV \cite{trends}. 

The bonding of both CO and CNH to transition metals is often discussed in terms of the
Blyholder picture \cite{Blyholder}.
Binding involves donation from the filled lone pair on the C into
the partially occupied Au s shell together with $2\pi^*$ back donation into the 
$2\pi^*$ molecular orbital due to interaction with gold $d_{xz}$ and $d_{yz}$. 
We probe this picture for the bonding of CNH on the flat surface
by analyzing the angular character of the local density of states
for the adsorbed molecule and the gold slab.
The projected density of states (PDOS) for the  monolayer adsorbed onto a 
Au(111) slab at 1/3 coverage in the atop configuration is plotted in Fig. 2 
along with the PDOS for a free CNH monolayer and a clean Au(111) slab. 
Also shown are isosurface plots of selected wavefunctions ($\Gamma$-point)
corresponding to certain energy regions in the PDOS of the Au(111)-CNH system.

In the PDOS of the free CNH monolayer several very sharp peaks are observed 
which can be associated with the molecular orbitals $4\sigma$, $1\pi$, $5\sigma$ and 
$2\pi^*$ (the last being unoccupied). 
After adsorption the $4\sigma$ level does not change much 
and there is no discernable interaction with gold.
This is not surprising considering that this level lies well below the bottom
of the gold valence band. 
In contrast, $1\pi$ and $5\sigma$ levels undergo substantial reorganization.
The $5\sigma$ state strongly interacts with 
hybridized $s$ and $d_{z^2}$ orbitals of gold.
A localized state is formed below the bottom of the gold valence band. 
In addition, there is a state of anti-bonding character that can be identified 
above the occupied gold d-band, but below the Fermi energy
(around -1 eV in Fig. 2).
These states are very similar to the hybrids that forms in the AuCNH radical.
In the radical, the anti-bonding state is half occupied.
As a result of the strong $\sigma$ interaction,
the $1\pi$ level now lies above $5\sigma$ derived level.
The $1\pi$ level interacts with gold $d_{xz}$ and $d_{yz}$ orbitals. 
This results in two peaks visible in the PDOS at
around $-7$ eV and $-5$ eV, of bonding and anti-bonding character respectively, 
as can be seen from the wavefunction plots.
The bonding combination appears to form a localized state just below the bottom
of the Au d-bands, while the anti-bonding combination is resonant with the d-bands.
Finally, the $2\pi^*$ peak from the CNH derived state density
extends somewhat below Fermi energy upon adsorption.
The corresponding wavefunction isosurface plot shows 
that these occupied states are of bonding character between gold p-like states available
near the Fermi energy and the CNH $2\pi^*$ orbital.
Hence back donation into the $2\pi^*$ orbital contributes to the bonding.
Above the Fermi energy, near the center of the resonance, the coupling
is weak, but clearly of anti-bonding character to gold $d_{xz}$ and $d_{yz}$ orbitals.
This qualitative picture from the PDOS is very similar to previous results
for CO adsorbed on the a-top site of gold (111) \cite{trends}.

It is hard to quantify this analysis of bonding character since the PDOS is calculated 
using only electron densities inside spheres surrounding the atoms.
However, another metric related to bonding trends is the change in C-N bond 
length upon adsorption on gold.
The calculation for the Au-CNH complex shows a C-N bond length of 1.173 $\AA$ in
the straight configuration.
Upon full relaxation to the bent configuration, 
the C-N bond is 1.198 $\AA$, compared to 1.176 $\AA$ in the free CNH.
This net weakening of the C-N bond upon bending traces to the pseudo Jahn-Teller effect 
driven by increased occupation of the antibonding 2$\pi^*$ orbital. 
On the other hand, upon adsorption on the Au(111) surface, 
the calculated C-N bond length is 
slightly decreased to 1.16 $\AA$.
This contrasts with results for CO adsorption, 
where the C-O bond length slightly increases
upon adsorption \cite{trends}.
The decrease in C-N bond length suggests slightly less $\pi$ back donation 
for CNH adsorption on the surface relative to the radical.
However, careful study of metal-carbonyl complexes 
reveals a more complex picture \cite{AuCO}.
The final C-N bond length (and the corresponding stretch frequency)
reflects the balance between $\sigma$ bond polarization due to the metal
and $\pi^*$ back donation effects.
This balance is affected by electrostatics, particularly in the limit of 
the positively charged carbonyl complex.
The positive charge near the C influences the heteropolar C-O
$\sigma$ bond, driving it to be more symmetrical and stronger.
This also drives an increase in the C-O stretch frequency in the carbonyls.
One would expect similar effects on the C-N bond for the CNH case.
This is consistent with the observed increase 
of the C-N stretch frequency in isocyanides adsorbed 
on gold \cite{7,8}.

The initial dipole on the molecule and the induced dipole upon adsorption
both play a role.
The CNH molecule has a substantial dipole moment, in contrast to the CO molecule.
Therefore, in the case of CNH, the binding 
energy contains some contribution from the polarization energy due to the interaction 
between the molecular dipole and the gold surface. 
Our estimates indicated a modest effect (0.1 eV),
which is however a non-trivial fraction of the final adsorption energy
for an isolated CNH on the flat surface (0.23 eV).
Furthermore, the net dipole inhibits formation of a dense film on the surface.
The inferred induced dipole on the metal surface
is substantially smaller than that found in the isolated Au-CNH radical.
This is indicative of 
less net charge transfer in comparison to the radical.  
This is not surprising from two points of view.
First, the work function of Au(111) (5.31 eV \cite{michaelson}) is smaller
than the chemical potential of the Au atom 
(conventionally the average of the ionization potential
and the electron affinity, 5.76 eV \cite{augas}).
This inhibits the sigma donation from the lone pair.
Second, the surface Au s-state is substantially involved in band formation, having nine
nearest neighbors.
This interferes with hybridization to the lone pair, reducing the net energy
gain for the sigma donation process.

The role of charge transfer has been controversial in the analysis of
these weakly adsorbed systems and the impact of the hybridization
implicit in the Blyholder picture has been debated
\cite{trends, ammonia, pyridine, bipyridine}.
Bilic {\it et al.} argue that the main mechanisms 
of binding of NH$_3$ to the gold surface are 
polarization effects and dispersive interactions, 
not the covalent bonding \cite{ammonia}. 
In their work on pyridine \cite{pyridine}, they find more evidence for 
charge transfer, but still argue that covalent effects are minimal.
A more recent study of pyridine binding to Au \cite{bipyridine} analyzed the impact of
Au coordination and suggested a more prominent role for hybridization effects.
The utility of the Blyholder picture becomes more apparent when examining
trends in the binding, e.g. for bonding to different metals \cite{trends}.
Alternatively, it gives a way to rationalize the trends for the binding
of different molecules to the same metal, as described for the Au radicals in Section II.
Finally, the sigma donation becomes more prominent for the case
of binding to an Au adatom on the surface. 
We found a large increase in binding energy of CNH compared to the flat surface (0.9 eV).
A similar increase was recently reported for NH$_3$ (0.4 eV) \cite{venkataraman}.
The donor-acceptor binding 
to the adatom is also enhanced due to the slight positive charge
on the Au adatom, similar to the increased binding energy for the Au(I) cation to
several ligands \cite{schroder}.

The small calculated value of adsorption energy of isocyanides on 
the flat gold surface seems to contradict the experimental observation of
SAM formation \cite{7}.
Two caveats are due here. 
First, layer formation takes place in a solution. 
Solvation effects can screen the dipole-dipole interactions 
and facilitate layer growth from nuclei.
Also, a layer of larger molecules (e.g. alkanes)
may be more stable than the binding energy of the
CNH link moiety would suggest due to attractive inter-molecular interactions 
between the extended molecules in the layer, e.g. due to dispersive interactions
between neighboring alkanes.
Second, the presence of undercoordinated Au binding sites at step edges
on the surface may facilitate nucleation of the isonitrile layers.
We have found that the binding energy to the Au adatom is substantially larger
than the binding energy to the top site on the flat surface.
This may also account for the binding to powdered gold samples and nanoparticles
where undercoordinated Au sites would be common.

Assessing the accuracy of DFT calculations for these weakly bound systems
is difficult.
The information for the Au radicals was already assessed in Sec. II.
In the case of 
Au-NH$_3$, the GGA Au to NH$_3$ binding energy
is smaller than accurate quantum chemistry calculations
by about 0.2 eV. 
Bilic {\it et al.} studied 
the binding of NH$_3$ to the gold surface 
using the GGA with the PW91 exchange-correlation functional,
which in this particular case gave an adsorption energy in 
quantitative agreement with experiment (within 0.1 eV) \cite{ammonia}. 
For the Au-CO radical, the binding energy is less well established, as discussed
in Sec. II.
Gajdo {\it et al.} studied adsorption of 
CO on (111) surface of various transition and noble metals \cite{trends}. 
In particular, they report results for CO adsorption 
on Au and Ag surfaces obtained with PW91 and 
a revised form of the PBE exchange-correlation potentials
which has in other cases given more accurate adsorption energies \cite{hammer}.
Adsorption energies were very sensitive to the choice of the 
functional.
Both functionals underestimate adsorption energies. 
The revised PBE leads to endothermic adsorption while PW91 underestimates
the binding energy by about 0.2 eV.
The essential mechanisms of binding are very similar for CO and CNH molecules, 
as it can be seen from the comparison of PDOS pictures. 
The experience with CO on the flat Au surface suggests that the
present DFT calculations may underestimate the binding energy
for the CNH case as well.
Also, in the CNH case, we have found that the LDA overestimates
the radical binding energies and predicts the wrong binding site on the flat surface.
The role of dispersion forces in the binding
has been debated.
The evidence just summarized points to the GGA calculations 
underestimating the binding energy of CO and CNH on the Au surface.
This is consistent with experience in the case of intermolecular interactions
for closed shell molecules which are dominated by dispersion forces.
The GGA approximations are known to underestimate 
binding in these cases while LDA over estimates them \cite{meijer}.

\begin{figure*}
\includegraphics{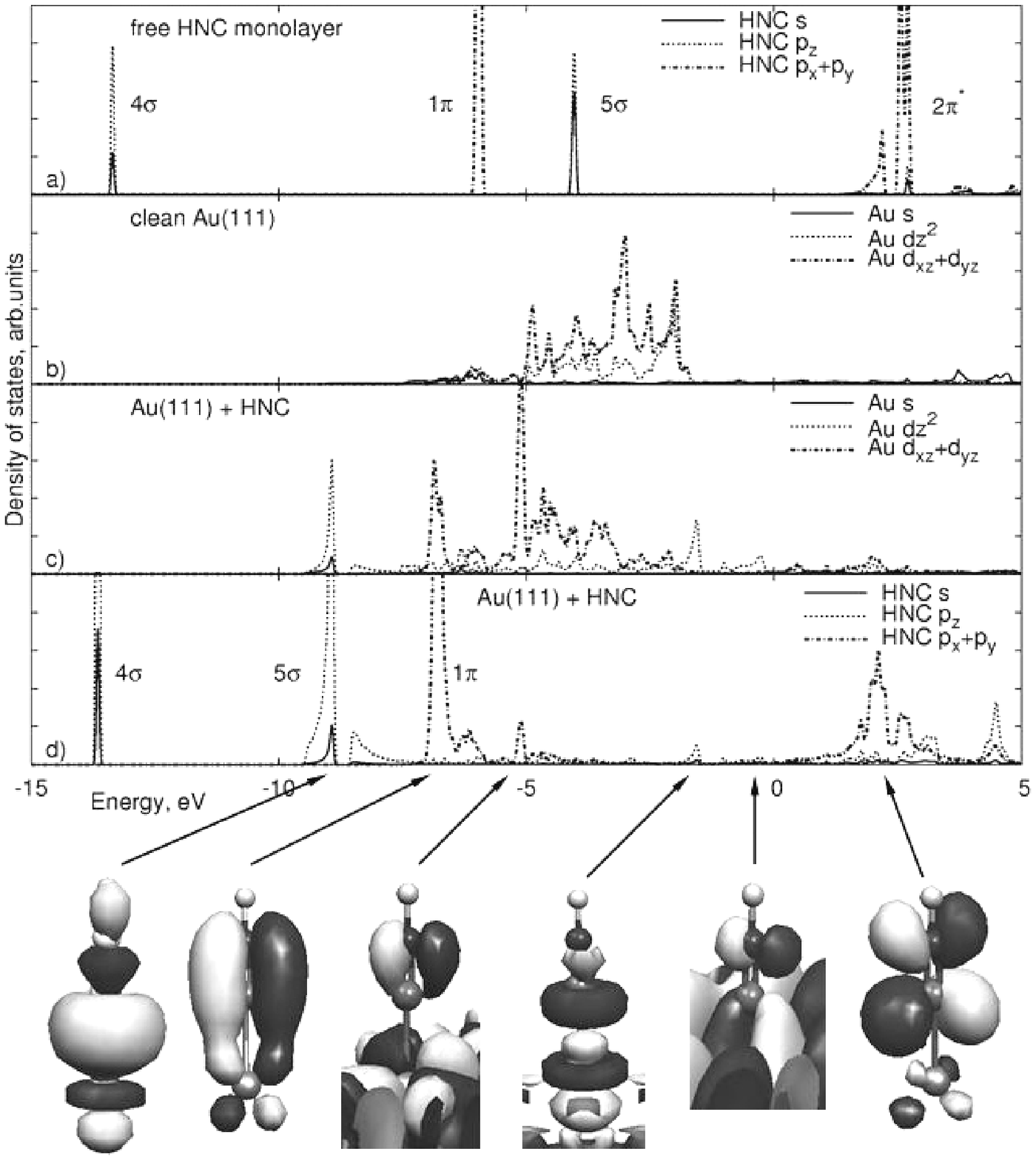}
\caption{Projected density of states onto: 
(a) the HNC molecule in free-standing monolayer; 
(b) the surface Au atom in the clean Au(111) slab; 
(c) the Au atom bound to the molecule in the Au(111)-CNH system (1/3 ML, atop site); 
(d) the HNC molecule in the Au(111)-CNH system. 
Panel (e) shows isosurface plots of selected wavefunctions (at $\Gamma$-point)
of the Au(111)-CNH system at energies indicated by the arrows.
}
\end{figure*}  

\noindent {\bf Acknowledgements}  We thank Tao Sun for help with the
WIEN2k calculations.  We thank J. Davenport for help and for time
on computers at the Computational Science Center at Brookhaven National Laboratory.  We thank K. K. Likharev
for use of Njal supercomputer cluster at Stony Brook.  We thank M. R. Pederson and T. Baruah for 
doing some useful molecular calculations. Work at Stony Brook was supported in part by
NSF Grant NIRT-0304122.  Work at BNL was supported by U.S. DOE under contract No. DEAC 02-98 CH 10886.
Work at Columbia University was supported by the 
Nanoscale Science and Engineering Initiative of the National Science Foundation 
under NSF Award Number CHE-0117752 and by the New York State Office of Science, 
Technology, and Academic Research (NYSTAR).


\begin{thebibliography}{99}

\bibitem{nuzzo} R.G. Nuzzo and D.L. Allara, J. Am. Chem. Soc. {\bf 105}, 
4481 (1983).

\bibitem{ulman} A. Ulman, Chem. Rev. {\bf 96}, 1533 (1996).

\bibitem{schreiber04} F. Schreiber, J. Phys.: Condens. Matter {\bf 16}, R881 (2004).

\bibitem{tour} L. Jones, II, J.S. Schumm, and J.M. Tour, J. Organic Chem. {\bf 62},
1388 (1997).

\bibitem{reed} M.A. Reed, C. Zhou, C.J. Muller, T.P. Burgin, and J.M. Tour,
Sci. {\bf 278}, 252 (1997).

\bibitem{salomon} A. Salomon, D. Cahen, S. Lindsay, J. Tomfohr, V.B. Engelkes,
and C.D. Frisbie, Adv. Mater. {\bf 15}, 1881 (2003).

\bibitem{basch} H. Basch, R. Cohen, and M.A. Ratner, Nano Lett. {\bf 5}, 1668 (2005).

\bibitem{poirrer} G.E. Poirer, Chem. Rev. {bf 97}, 1117 (1997).

\bibitem{schreiber00} F. Schreiber, Prog. Surf. Sci. {\bf 65}, 151 (2000).

\bibitem{fischer} D. Fischer, A. Curioni, and W. Andreoni, Lang. {\bf 19}, 3567 (2003).

\bibitem{nuckolls} G.S. Tulevski, M.B. Myers, M.S. Hybertsen, M.L. Steigerwald,
and C. Nuckolls, Sci. {\bf 309}, 591 (2005).

\bibitem{mcbreen} M. Siaj and P.H. McBreen, Sci. {\bf 309}, 588 (2005).

\bibitem{venkataraman} L. Venkataraman, J.E. Klare, I.W. Tam, 
C. Nuckolls, M.S Hybertsen and M. Steigerwald, Nano Lett. {\bf 5}, 458 (2006).

\bibitem{lin} S. Lin and R.L. McCarley, Lang. {\bf 15}, 151 (1999).

\bibitem{8} M. J Robertson, R. J. Angelici, Langmuir {\bf 10}, 1488, (1994).

\bibitem{9} K.Shih, R. J. Angelici, Langmuir {\bf 11}, 2539 (1995).

\bibitem{10} S.J. Bae, Ch. Lee, I.S. Choi, Ch.-S. Hwang, M. Gong, K. Kim, 
and S.-W. Joo, J. Phys. Chem. B {\bf 106}, 7076 (2002).

\bibitem{11} H. S. Kim, S. J. Lee, N. H. Kim, J. K. Yoon, H. K. Park, 
and K. Kim, Langmuir {\bf 19}, 6701 (2003).

\bibitem{7} J. I. Henderson, S. Feng, T. Bein, and C. P. Kubiak, 
Langmuir {\bf 16}, 6183 (2000).

\bibitem{13} J. M. Seminario, A.G. Zacarias, and J.M. Tour, J. Am. Chem. Soc.
{\bf 121}, 411 (1999).

\bibitem{1} J. Chen, L.C. Calvet, M.A. Reed, D.W. Carr, D.S. Grubisha, 
D.W. Bennett, Chem. Phys. Lett. {\bf 313}, 741 (1999).

\bibitem{2} J.-O Lee, G. Lientschnig, F. Wiertz, M. Struijk, R.A.J. Janssen, 
R. Egberink, D.N. Reinhoudt, P. Hadley, and C. Dekker, 
Nano Letters, {\bf 3}, 113 (2003).

\bibitem{3} Y. Xue and M. A. Ratner, Phys. Rev. B {\bf 69}, 085403 (2004).

\bibitem{trends} M. Gajdos, A. Eichler and J. Hafner,
J. Phys. Condens. Matter  {\bf 16}, 1141 (2004).

\bibitem{ammonia} A. Bilic, J. R. Reimers, N. S. Hush, and J. Hafner, J. Chem. Phys. {\bf 116}, 8981 (2002)

\bibitem{Blyholder} G. Blyholder, J. Phys. Chem. {\bf 68}, 2772 (1964). 

\bibitem{PBE} J. P. Perdew, K. Burke, and M. Ernzerhof,
Phys. Rev. Letters {\bf 77}, 3865 (1996).

\bibitem{hammer} B. Hammer, L.B. Hansen, and J.K. Norskov, Phys. Rev. B
{\bf 59}, 7413 (1999).

\bibitem{PBEtest} S. Kurth, J. P. Perdew, P. Blaha, Int. J. of Quant. Chem. {\bf 75}, 889 (1999).

\bibitem{feibelman} P. J. Feibelman, B. Hammer, J. K. Norskov, F. Wagner, M. Scheffler, R. Stumpf, R. Watwe, and J. Dumesic, J. Phys. Chem. B {\bf 105}, 4018 (2001).

\bibitem{gross} A. Gross, Surf. Sci. Rep. {\bf 32}, 291 (1998).

\bibitem{LDA} 
The parameterization of Ref. \cite{goedecker} is used with 
the Abinit slab calculations. Perdew and Wang parametrization \cite{PW92} is used in NRLMOL and WIEN2k 
calculations.

\bibitem{PW92}  J.P. Perdew and Y. Wang, Phys.Rev. B {\bf 45}, 13244 (1992).

\bibitem{goedecker} S. Goedecker, M. Teter, and J. Hutter, Phys. Rev. B
{\bf 54}, 1703 (1996).

\bibitem{NRLMOL} M.R. Pederson and K.A. Jackson, Phys. Rev. B {\bf 41}, 7453
(1990); K.A. Jackson and M.R. Pederson, Phys. Rev. B
{\bf 42}, 3276 (1991).

\bibitem{NRLMOLpsp}  D. Porezag and M.R. Pederson, Phys. Stat. Solidi (b) {\bf 217}, 219 (2000).

\bibitem{NRLMOLbasis}  D. Porezag and M.R. Pederson, Phys. Rev. A, {\bf 60}, 2840 (1999); D. V. Porezag, PhD thesis: http://archiv.tu-hemnitz.de/pub/1997/0025.

\bibitem{12} P. Blaha, K. Schwarz, G. K. H. Madsen, D. Kvasnicka and J. Luitz, 
WIEN2k, An Augmented Plane Wave + Local Orbitals Program for Calculating 
Crystal Properties (Karlheinz Schwarz, Techn. Universitat Wien, Austria), 
2001. ISBN 3-9501031-1-2

\bibitem{abinit} The ABINIT code is a common project of the Université Catholique de Louvain, C. I., and other contributors. URL http://www.abinit.org.

\bibitem{gonze} X. Gonze, J.M. Beuken, R. Caracas, F. Detraux, M. Fuchs,  
G.M. Rignanese, L. Sindic, M. Verstraete, G. Zerah, F. Jollet, M. Torrent, 
A. Roy, M. Mikami, P. Ghosez, J.Y. Raty, and D.C. Allan, 
Computational Materials Science {\bf 25}, 478 (2002).

\bibitem{HGH} C. Hartwigsen, S. Goedecker, and J. Hutter, Phys. Rev. B { \bf 58}, 3641 (1998). 

\bibitem{TM} N. Troullier and J.L. Martins, Phys. Rev. B { \bf 43}, 8861 (1991).

\bibitem{FHI} M. Fuchs and M. Scheffler, Computer Physics Communications { \bf 119}, 67 (1999).

\bibitem{tetrahedron} P. E. Blochl, O. Jepsen and O. K. Andersen, Phys. Rev B {\bf 49}, 16223 (1994).

\bibitem{lambropoulos} N.A. Lambropoulos, J.R. Reimers, and N.S. Hush, J. Chem. Phys.
{\bf 116}, 10277 (2002).

\bibitem{dargel} T.K. Dargel, R.H. Hertwig, W. Koch, and H. Horn, 
J. Chem. Phys. {\bf 108}, 3876 (1998).

\bibitem{schwerdtfeger} P. Schwerdtfeger and G.A. Bowmaker, J. Chem. Phys.
{\bf 100}, 4487 (1994).

\bibitem{liang} B. Liang and L. Andrews, J. Phys. Chem. A
{\bf 104}, 9156 (2000).

\bibitem{mendizabal} F. Mendizabal, Organometallics
{\bf 20}, 261 (2001).

\bibitem{wu} X. Wu, L. Senapati, S.K. Nayak, A. Selloni, and M. Hajaligol,
J. Chem. Phys. {\bf 117}, 4010 (2002).

\bibitem{schroder} D. Schroder, H. Schwarz, J. Hrusak, and P. Pyykko,
Inorg. Chem., {\bf 37}, 624 (1998).

\bibitem{aminecapped} D. V. Leff, L. Brandt, and J. R. Heath, Langmuir {\bf 12}, 4723 (1996).

\bibitem{phosphinecapped} W. W. Weare, S. M. Reed, M. G. Warner, and J. E. Hutchison, J. Am. Chem. Soc. {\bf 122}, 12890 (2000). 

\bibitem{AuCO}  A.J. Lupinetti, S. Fau, G. Frenking, and S.H.Strauss, J. Phys. Chem. A {\bf 101}, 9551 (1997).

\bibitem{michaelson} H. B. Michaelson, J. Appl. Phys. {\bf 48}, 4729 (1977).

\bibitem{augas} J. M. Dyke, N. K. Fayad, A. Morris, and I. R. Trickle,
J. Phys. B {\bf 12}. 2985 (1979); 
G. Gantefor, S. Kraus, and W. Eberhardt, J. Electron Spectrosc. Relat.
phenom. {\bf 88}, 35 (1998).

\bibitem{pyridine} A. Bilic, J.R. Reimers, and N.S. Hush, 
J. Phys. Chem. B {\bf 106}, 6740, (2002).

\bibitem{bipyridine} R. Stadler, K.S. Thygesen, and K.W. Jacobsen, Phys. Rev. B 
{\bf 72}, 241401(R) (2005).

\bibitem{meijer} E.J. Meijer and M. Sprik, J. Chem. Phys.
{\bf 105}, 8684 (1996).

\end{thebibliography}
\end{document}